# Intense Dark Exciton Emission from Strongly Quantum Confined $CsPbBr_3$ Nanocrystals

**Authors:** Daniel Rossi[1], Xiaohan Liu[3], Yangjin Lee[1,5], Mohit Khurana[3], Joseph Puthenpurayil[2], Kwanpyo Kim[1,5], Alexey Akimov[3,6,7], Jinwoo Cheon[1,4] * and Dong Hee Son[2,1] *

**Affiliations:**
[1]Center for Nanomedicine, Institute for Basic Science (IBS), Seoul 03722, Republic of Korea
[2]Department of Chemistry, Texas A&M University, College Station, Texas, 777843, USA
[3]Department of Physics, Texas A&M University, College Station, Texas, 777843, USA
[4]Department of Chemistry, Yonsei University, Seoul 03722, Republic of Korea
[5]Department of Physics, Yonsei University, Seoul 03722, Republic of Korea
[6]Russian Quantum Center, Skolkovo, Moscow, 143025, Russia
[7]PN Lebedev Institute RAS, Moscow, 119991, Russia

*Correspondence to:
E-mail: dhson@chem.tamu.edu
E-mail: jcheon@yonsei.ac.kr

**Abstract**

Dark exciton as the lowest-energy (ground) exciton state in metal halide perovskite nanocrystals is a subject of much interest. This is because the superior performance of perovskites as the photon source combined with long lifetime of dark exciton is very attractive for the applications such as quantum information processing. However, the direct observation of the long-lived dark exciton emission confirming the accessibility to dark ground exciton state has remained elusive. Here, we report the intense photoluminescence from dark exciton with microsecond lifetime in strongly confined $CsPbBr_3$ nanocrystals and reveal the crucial role of confinement in accessing the dark ground exciton state. This study establishes the potential of strongly quantum confined perovskite nanostructures as the excellent platform to harvest the benefits of extremely long-lived dark exciton.

Metal halide perovskite (MHP) nanocrystals (NCs) have gained explosive interest as a superior source of photons and charges with high luminescence quantum yield and long carrier diffusion length compared to many other semiconductor NCs.(*1, 2*) The integration of MHP NCs into the technological applications, such as solar cells and light emitting devices has driven intensive research on characterization and structural control of the properties of excitons.(*3-7*). Recently, the exciton fine structure of MHP NCs, in particular regarding the access to the dark exciton state, has become a subject of much interest, which has broad applications that can benefit from its extremely long lifetime such as quantum information processing. (*8-12*)

The optically inactive dark exciton appears as the lowest-energy (ground) exciton state in most semiconductor NCs, therefore, the dark exciton state is populated through the transition from the initially excited bright exciton to dark exciton state. At low temperatures with thermal energy ($kT$) smaller than the bright-dark energy splitting ($\Delta E_{BD}$), a substantial dark exciton population can be obtained giving rise to long-lived photoluminescence (PL) from the radiative relaxation of dark excitons. (*13-15*) Recently, the reversal of the bright and dark exciton level ordering was reported in cesium lead halide ($CsPbX_3$) NCs, based on the observation of only bright exciton PL in single-particle studies at cryogenic temperatures, indicating the loss of the typical pathway to reach the dark exciton level.(*8*) On the other hand, a study in $FAPbBr_3$ NCs under magnetic field reported the signature of a dark exciton ~2.5 meV below the bright exciton level in the single-particle PL spectra.(*9*) Another study in Mn-doped $CsPbCl_3$ NCs also argued for the dark ground state exciton from the observation of the weak but longer-lived decay component in the lower-temperature PL.(*10*) However, because of the dominance of the bright exciton PL even at cryogenic temperatures, and the absence of a direct measurement of the dark exciton relaxation, the access to dark exciton as well, as its energetic and dynamic characteristics, still remain elusive in this new class of semiconductor NCs.

Here, we report the direct observation of intense and long-lived PL with 1-10 μs lifetime from dark ground exciton state in *strongly quantum confined* $CsPbBr_3$ NCs revealing the importance of confinement-enhanced $\Delta E_{BD}$ in gaining a facile access to dark exciton. In contrast to MHP NCs studied earlier that were in the weak confinement regime, strongly quantum confined $CsPbBr_3$ NCs at cryogenic temperatures exhibit the PL primarily from dark exciton. The bright-dark energy splitting ($\Delta E_{BD}$), determined directly from the PL spectra, is nearly an order of magnitude larger than weakly confined $FAPbBr_3$ NCs and strongly confined NCs of many other semiconductors.(*9, 13, 16*) The larger $\Delta E_{BD}$ in strongly confined $CsPbBr_3$ NCs enables the access to the dark exciton at higher temperatures, which is particularly important for the applications of dark exciton. Furthermore, intense dark exciton PL was observed regardless of the dimensionality of the confinement, making all strongly confined 0D (quantum dot), 1D (nanowire) and 2D (nanoplatelets) NCs viable nanostructures to exploit dark excitons longevity. The results from this study demonstrate the potential of strongly quantum confined MHP NCs as the excellent material platform to utilize dark exciton, combining the benefits of the superb properties of MHP as the source of photons and charges with large confinement-enhanced $\Delta E_{BD}$.

Fig. 1 shows the electron microscopy images and room-temperature optical spectra of weakly confined NCs and strongly confined quantum dots (QDs), nanowires (NWs), and nanoplatelets (NPLs) of $CsPbBr_3$ used to examine the PL from dark exciton. The size of the NCs in the quantum confined dimension indicated in Fig. 1 for the QDs, NWs, and NPLs is much smaller than twice the exciton Bohr radius of $CsPbBr_3$ ($2a_B$ = 7 nm) imposing strong quantum confinement in varying dimensionality.(*1*) The weakly confined NCs were prepared following the procedure reported by Protesescu et al.(*1*) Strongly confined QDs, NWs, and NPLs were synthesized following the methods developed by Dong et al. which can control the size and morphology precisely with high ensemble uniformity, as reflected in the microscopy images and the optical spectra showing the well-defined features of strongly confined excitons.(*7, 17*) For the optical measurements, all the samples were passivated with dimethyl-dioctadecyl-ammonium bromide (DDAB) via ligand exchange for improved stability and PL quantum yield in polystyrene (PS) matrix used to disperse the NCs for PL measurements. Detailed descriptions of the synthesis and measurement methods are in the supplementary information.

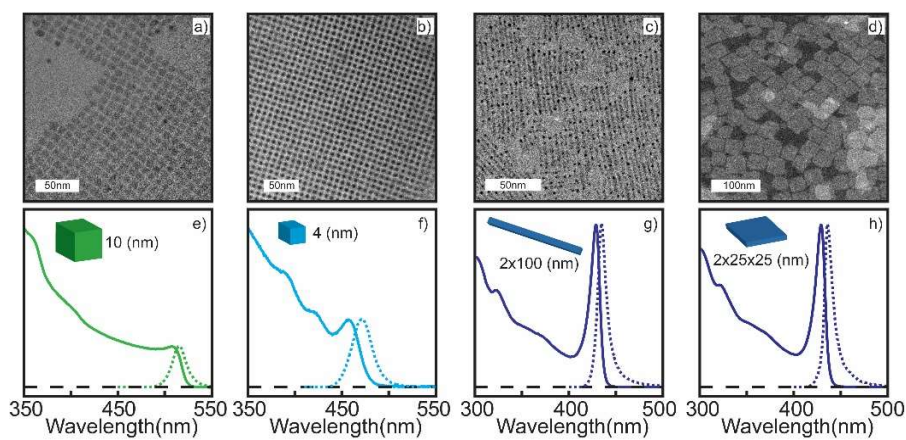

**Fig 1:** Electron microscopy images (a-d), and optical spectra (e-h) of weakly confined NCs, and strongly confined QDs, NWs, and NPLs (left to right). The cartoon depicting the NC morphology and the size are indicated in each panel. Solid and dashed lines in the spectra are for the absorption and photoluminescence respectively. (a-c) are TEM images and (d) is STEM image.

Fig. 2 shows the time-resolved PL spectra (a-d), spectrally integrated PL decay dynamics (e-h) and steady state PL spectra (i-l) following above-bandgap excitation at 5K for the four $CsPbBr_3$ NC samples shown in Fig. 1. While the PL spectra at 300K exhibits a single well-defined PL feature which decays on the several ns time scale for all samples, (Fig. S1-S5), at 5K the PL shows very different spectral and dynamic features. In the weakly confined $CsPbBr_3$ NCs, a single PL peak with ~500 ps decay time is observed (Fig. 2a, e, i), consistent with the results from earlier studies.(*8, 11*) Because of the intense and fast-decaying PL on the sub-ns time scale even at 5K, the earlier study concluded that the dark exciton level is above the bright exciton level in $CsPbBr_3$ NCs.(*8*)

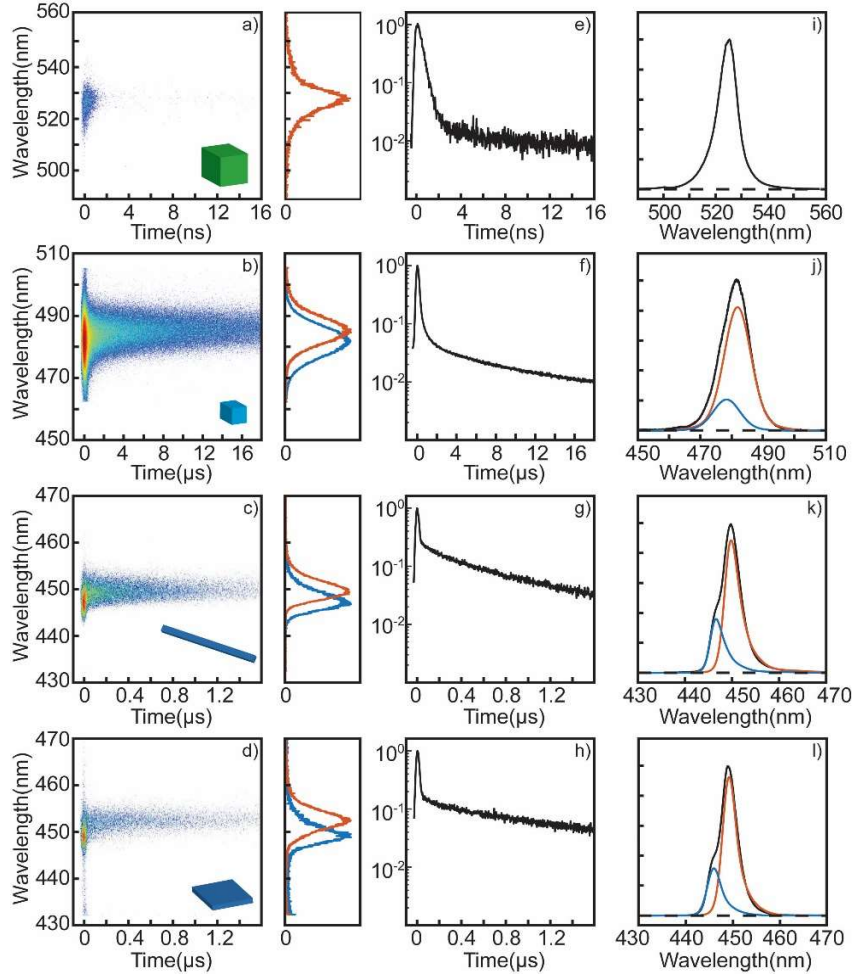

**Fig 2:** Time-resolved PL spectra (a-d), normalized spectrally integrated PL decay dynamics (e-h), and steady state PL spectra (i-l) of four different CsPbBr$_3$ NCs measured at 5K. The data are for weakly confined NCs, QDs, NWs, and NPLs from top to bottom. The two spectra shown next to panels b-d are the time-gated PL spectra taken near 0 ps (blue) and 0.5 μs (red). These two spectra are used to fit the steady state spectra in (i-l).

| Sample | $\Delta E_{BD}$ (meV) | $\tau_{slow}=\tau_D$ (μs) | $\tau_{fast}$ (ps) | $\tau_B$ (ps) | $\tau_{BD}$ (ps) | $I_D/I_{tot}$ (%) |
|---|---|---|---|---|---|---|
| QD | 17 | 10 | 387 | 1600 | 510 | 76 |
| NW | 19 | 0.85 | 44 | 160 | 62 | 68 |
| NPL | 22 | 1.2 | 37 | 124 | 54 | 73 |

**Table1:** Parameters extracted from the PL decay data at 5K. $\Delta E_{BD}$ and $\tau_D$ were extracted directly from the time-resolved PL spectra and PL decay kinetics. The fraction of dark exciton PL from total PL was obtained by fitting of the steady state PL spectra. $\tau_B$ and $\tau_{BD}$ were estimated from the analysis of the steady state PL intensity and $\tau_{fast}$. Accurate $\tau_{fast}$ was obtained from short-time window PL decay kinetics data (Fig. S8) that provide higher time resolution as detailed in supplementary information.

In contrast, all three strongly confined NCs of varying dimensionalities, i.e. QDs, NWs and NPLs, exhibit two different PL peaks with decay time constants that differ by many orders of magnitude (Fig. 2 b-d). Time gating the PL spectra near 0 ps and 0.5 μs readily reveals the two PL peaks separated by 17-22 meV as shown to the right of Fig. 2b-d. The fast and slow components of the PL decay kinetics in Fig. 2 f-h with corresponding time constants of $\tau_{fast}$ and $\tau_{slow}$ are associated with the higher-energy and lower-energy PL peaks respectively. Fitting the steady state PL spectra (Fig. 2 j-l) with the two PL peaks indicates that the lower-energy PL constitutes ~70 % of the total photons emitted at 5K in these NCs.

We assign the higher-energy and lower-energy PL to the bright and dark excitons respectively, as will be discussed in detail shortly. Since the bright and dark exciton PL are clearly separable in the time-resolved spectra, $\Delta E_{BD}$ is directly determined from the separation between the two PL peaks. Table 1 summarizes $\Delta E_{BD}$, $\tau_{fast}$ and $\tau_{slow}$ of the PL decay kinetics, and the fraction of photons from the dark exciton in the PL at 5K. Since $\Delta E_{BD}$ is much larger than $kT$ at 5K, prohibiting thermal excitation from the dark to bright exciton states, $\tau_{slow}$ can reliably be taken as the dark exciton lifetime $\tau_D$. On the other hand, $\tau_{fast}$ reflects the decay of the bright exciton population via radiative and nonradiative relaxation combined with transition from the bright to the dark exciton. The estimation of the bright exciton lifetime ($\tau_B$) and the time constant for the bright-to-dark transition ($\tau_{BD}$) has been made from the analysis of the relative intensities of the bright and dark exciton PL at 5K with a few assumptions, as described in detail in the supplementary information.

In Table 1, $\tau_{BD}$ is significantly shorter than $\tau_B$ for all three strongly confined NCs, indicating that dark exciton can be reached effectively from the initially excited bright exciton. Therefore, the stark contrast of the PL dynamics between the weakly and strongly confined NCs at cryogenic temperatures is mainly due to the large $\Delta E_{BD}$ enhanced by confinement combined with efficient bright-to-dark transition. The confinement enhancement of $\Delta E_{BD}$ predicted earlier from theoretical study is corroborated by the comparison of the $\Delta E_{BD}$ of QDs and NPLs with varying degree of quantum confinement at 5K. (*18, 19*) Fig. S6 (QDs) and S7 (NPLs) comparing the lower and higher-energy PL peaks from the time-resolved PL spectra at 5K clearly show the larger $\Delta E_{BD}$ for the stronger confinement. While it was conjectured that slow bright-to-dark transition may be partially responsible for the difficulty in observing dark exciton PL in the MHP NCs with dark ground exciton state, the present study suggests that the strongly size-dependent $\Delta E_{BD}$ is a more critical factor that determines the accessibility to the dark exciton. For NWs and NPLs, the major contribution to the PL was from dark exciton even at ~20 K (Fig. S2-S5), due to the large $\Delta E_{BD}$, which is much larger than in other strongly confined semiconductor NCs, such as CdSe QDs (1.7 meV) and CdSe NPLs (4.5 meV).(*16, 20*) Interestingly, $\tau_B$ of NWs (160 ps) and NPLs (124 ps) are nearly an order of magnitude shorter than in QDs. The faster $\tau_B$ in NWs and NPLs compared to QDs is reminiscent of the enhanced exciton recombination rate observed in other 1D and 2D confined semiconductors, reflecting the larger exciton binding energies and enhanced electron-hole correlations.(*16, 21*) Despite the rapid relaxation of bright exciton in NWs and NPLs, even more rapid bright-to-dark transition enables reaching dark exciton as efficiently as in QDs.

The assignment of μs-lived lower-energy PL to the dark exciton has been made as follows by ruling out other processes that can produce a PL redshifted from the bright exciton PL in MHP NCs and performing the magneto fluorescence measurements. Self-trapped exciton, phonon replica, and defect emission have been discussed previously as the origin of redshifted PL observed in several MHP materials, both in bulk and nanocrystalline forms.(*22-26*) In addition, trion emission and interparticle excimer-like emission have been discussed as the origin of certain PL features redshifted from the exciton PL in CdSe NPLs, although such observations have not been reported for MHP NCs yet.(27, 28) These alternative possibilities were ruled out by examining the difference in the spectral characteristics of the PL, such as the linewidth, lifetime, the magnitude of redshift from the bright exciton, and the dependence of the lifetime on external magnetic field.

Self-trapped exciton has been discussed extensively to explain a broad PL feature redshifted from the exciton PL observed in some MHP NCs, prominent under sub-gap excitation condition.(22) PL from the self-trapped exciton typically has a much larger spectral linewidth and Stokes shift than the exciton PL due to the larger lattice displacement associated with these transitions. In $CsPbBr_3$ NCs, the PL from self-trapped exciton was observed ~100 meV below the exciton PL, with three times broader linewidth and a lifetime of ~170 ns, very different from the lower-energy PL observed in the strongly confined $CsPbBr_3$ NCs of this study. Furthermore, the disappearance of the self-trapped exciton PL with above-bandgap excitation further rules out self-trapped exciton as the origin for the μs-lived lower-energy PL observed in our study. Similarly, phonon overtones are observed in the low-temperature PL spectra for all semiconductors. The first phonon overtone will appear as a PL sideband redshifted by one optical phonon energy, with an intensity proportional to the exciton-phonon coupling. While $\Delta E_{BD}$ in this study is similar in the order of magnitude to the optical phonon energy of bulk $CsPbBr_3$, the large disparity between the lifetimes of the two PL peaks by many orders of magnitude immediately dismisses this possibility.

A redshifted PL feature with similar spectral linewidth to the exciton PL is often observed in the low-temperature PL spectra of CdSe NPLs. While its origin is still debated, two recent works suggested it may result from the excimer-like emission from stacked NPLs or trion emission.(27, 28) Considering that we observed the redshifted PL universally in all strongly quantum confined NCs of different dimensionalities, the excimer-like emission observed in CdSe NPLs is not likely to explain our observation in MHP NCs. Furthermore, the independence of the PL lifetime on the magnetic field from these two origins contrasts with the strongly magnetic field-dependent lifetime of the slow-decay component in the PL shown in Fig. 3, further ruling out these two options.

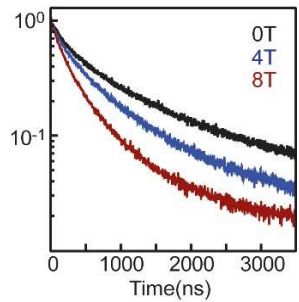

**Fig 3**: Magnetic field-dependent PL decay dynamics of $CsPbBr_3$ NPLs at liquid helium temperature. Only the slow-decay component of the PL from $CsPbBr_3$ NPLs is plotted.

Fig. 3 shows the dependence of the slow-decay component of the PL from $CsPbBr_3$ NPLs under 0T, 4T, and 8T magnetic field at liquid helium temperature. The lifetime shortens with increasing magnetic field, consistent with the description of the dark and bright exciton levels mixed by the magnetic field. The similar magnetic field dependence of the dark exciton lifetime was observed in recent studies on single FAPbBr$_3$ NCs and Mn-doped CsPbCl$_3$ NC ensembles.(*9, 10*) For self-trapped exciton, defect emission, trion emission and excimer-like emission, the dependence of the PL lifetime on the magnetic field is not expected, as has been experimentally confirmed in several studies.(*27, 28*) In addition to the discussion above, the dependence of the PL lifetime on the magnetic field supports decisively the assignment of the lower-energy PL to the dark exciton.

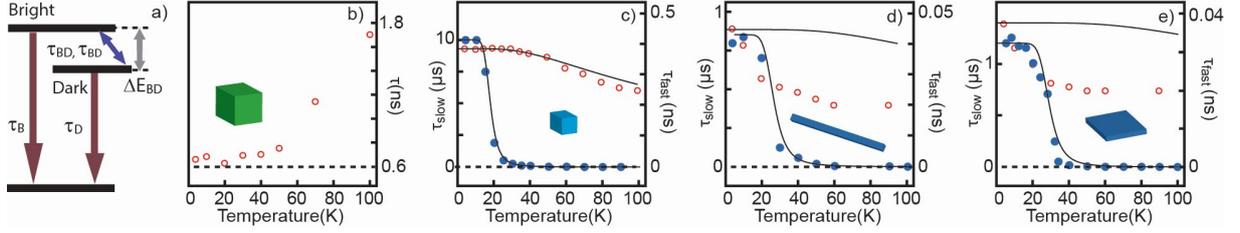

**Fig 4:** Kinetic model (a) used to analyze the temperature dependence of $\tau_{fast}$ (○) and $\tau_{slow}$ (●) from QDs (c), NWs (d) and NPLs (e). Curves are the values calculated using the kinetic model and nonadjustable parameters in Table 1. For weakly confined NCs (b), τ is from the single exponential fit of the PL decay data.

In order to gain further insights into the competitive dynamics of the exciton relaxation and bright-to-dark transition and to check the validity of the parameters in Table 1, the temperature-dependent $\tau_{fast}$ and $\tau_{slow}$ of the strongly confined NCs were analyzed with a kinetic model shown in Fig. 4a. Fig. 4 b-e compare $\tau_{fast}$ and $\tau_{slow}$ of QDs, NWs and NPLs with the single-exponential PL decay time of weakly confined NCs in the temperature range of 5-100 K. Under this kinetic model, $\tau_{fast}$ and $\tau_{slow}$ at each temperature are determined by the competition of the relaxation of bright and dark excitons separated by $\Delta E_{BD}$ and the temperature-dependent reversible transition between bright and dark states as explained in detail in supplementary information.(*29*) For the weakly confined NCs (Fig. 4b), the exciton PL decays largely single exponentially with decreasing time constant with decreasing temperature, similar to the earlier observation.(*8*) However, slight departure from single exponential decay is also observed between 5-50 K.(Fig. S1) For QDs, the temperature dependence of $\tau_{fast}$ and $\tau_{slow}$ is very well reproduced with the kinetic model using $\Delta E_{BD}$, $\tau_B$, $\tau_D$, and $\tau_{BD}$ in Table 1 as nonadjustable parameters, supporting the quantitative validity of the time constants and $\Delta E_{BD}$ extracted from the time-resolved spectra at 5K. The same analysis for NWs and NPLs reproduces $\tau_{slow}$ quite well, while $\tau_{fast}$ shows more discrepancy between the measured and calculated values. Rapid change of the measured $\tau_{fast}$ in 5-20 K suggests the possible involvement of an additional nonradiative decay channel of the bright exciton in NWs and NPLs not included in the kinetic model, since $\tau_{fast}$ is sensitive to $\tau_B$ and $\tau_{BD}$ in particular, consistent with the more complicated temperature dependence of the relative PL QY for those samples.(Fig S9)

Our results present clear evidence for the readily accessible dark ground exciton state in strongly quantum confined CsPbBr$_3$ NCs. In contrast to weakly confined NCs, strongly confined QDs, NWs and NPLs of CsPbBr$_3$ exhibit intense and long-lived dark exciton PL < ~20 K due to the large confinement-enhanced splitting between bright and dark states. This result establishes the exciting possibility to take advantage of the superior photonic properties of strongly confined MHPs in the applications utilizing long-lived dark states such as in quantum information processing.

**Funding:** DHS and AA acknowledge the National Science Foundation (CHE-1836538) for financial support. JC acknowledges Institute for Basic Science (IBS-R026-D1) for financial support.

**Author contributions:** DHS conceived the idea for the study. DR synthesized the materials and performed the low temperature streak measurements. XL and MK performed the low temperature magnetic field measurements under the supervision of AA. YL performed the TEM measurement under the supervision of KK and JC. DR and DHS wrote the manuscript. All authors discussed the results and commented on the manuscript. DHS and JC supervised the project.

**Competing interests:** None declared.

**Materials and Correspondence:** Correspondence should be addressed to Dong Hee Son and Jinwoo Cheon.

**Data and materials availability:** All (other) data needed to evaluate the conclusions in the paper are present in the paper or the supplementary information.

# Supplementary Information

## Intense Dark Exciton Emission from Strongly Quantum Confined CsPbBr$_3$ Nanocrystals

**Authors:** Daniel Rossi[1], Xiaohan Liu[3], Yangjin Lee[1,5], Mohit Khurana[3], Joseph Puthenpurayil[2], Kwanpyo Kim[1,5], Alexey Akimov[3,6,7], Jinwoo Cheon[1,4] * and Dong Hee Son[2,1] *

**Affiliations:**
[1]Center for Nanomedicine, Institute for Basic Science (IBS), Seoul 03722, Republic of Korea
[2]Department of Chemistry, Texas A&M University, College Station, Texas, 777843, USA
[3]Department of Physics, Texas A&M University, College Station, Texas, 777843, USA
[4]Department of Chemistry, Yonsei University, Seoul 03722, Republic of Korea
[5]Department of Physics, Yonsei University, Seoul 03722, Republic of Korea
[6]Russian Quantum Center, Skolkovo, Moscow, 143025, Russia
[7]PN Lebedev Institute RAS, Moscow, 119991, Russia

*Correspondence to:
E-mail: dhson@chem.tamu.edu
E-mail: jcheon@yonsei.ac.kr

**List of Supplementary information:**

**S1. Materials**
**S2. Preparation of CsPbBr$_3$ Nanocrystals**
**S3. Sample Characterization**
**S4. Temperature Dependent PL dynamics (Fig. S1-S5)**
**S5 Effects of Quantum Confinement on ΔE$_{BD}$ (Fig. S6-S7)**
**S6. Estimation of $\tau_B$ and $\tau_{BD}$ from the Steady State PL Spectra (Fig. S8-S9)**
**S7. Three-Level Kinetic Model and Temperature-Dependent $\tau_{fast}$ and $\tau_{slow}$ (Fig. S10)**

## S1. Materials

Materials: Cesium Carbonate ($Cs_2CO_3$, 99.995%, Trace Metal Basis, Sigma Aldrich), Lead Bromide ($PbBr_2$, 99.999%, Trace Metal Basis, Sigma Aldrich), Zinc Bromide ($ZnBr_2$, 99.999%, Trace Metal Basis, Sigma Aldrich), Cobalt Bromide ($CoBr_2$, 99%, Sigma Aldrich), Copper Bromide ($CuBr_2$, 99%, Sigma Aldrich), Oleylamine (OAm, 70%, Technical Grade, Sigma Aldrich), Oleic Acid(OA, 90%, Technical Grade, Sigma Aldrich), Octadecene (ODE, 90%, Technical Grade, Sigma Aldrich), Polystyrene(Sigma Aldrich, 192,000 average molecular weight).

## S2. Preparation of $CsPbBr_3$ Nanocrystals

Preparation of Cs Precursor Solution
All NC samples were prepared using the same $Cs_2CO_3$ precursor solution. 250 mg $Cs_2CO_3$, 1 mL OA, 7 mL ODE were loaded into a 50 mL three neck flask and vacuumed on a Schlenk line for 10 minutes at room temperature. The flask was refilled with Ar gas and heated to 150 ˚C. The solution was held at 150 ˚C until use.

Preparation of $CsPbBr_3$ weakly confined nanocrystals (NCs)
$CsPbBr_3$ NCs were prepared following the methods Protesescu et al.(*12*) Typically 70 mg $PbBr_2$, 5 mL ODE, 0.5 mL OA, and 0.5 mL OAm were loaded into a three neck flask and vacuumed on a Schlenk line at room temperature for 10 minutes. The solution was heated to 110 ˚C under vacuum and the flask was refilled with Ar gas. The solution as heated to 200 ˚C and held for 5 minutes. 0.4 mL of Cs precursor was injected into the reaction mixture and the reaction was quenched in an ice bath. The resulting NC suspension was centrifuged at 7000 rpm for 10 minutes and the supernatant was discarded. The precipitate NCs were dispersed in hexane and the solution was centrifuged at 17000RPM for 10 minutes. The supernatant was collected and stored under Ar at 4 ˚C until use.

Preparation of $CsPbBr_3$ strongly confined quantum dots (QDs)
Strongly confined $CsPbBr_3$ QDs (4nm) were prepared via methods recently developed by Dong et al.(*18*) In a glove box, 600 mg $ZnBr_2$ 250 mg $PbBr_2$ 8 mL ODE, 4 mL OA and 4 mL OAm were loaded into a 25 mL 3 neck flask and transferred to a Schlenk line where they were vacuumed at 100 ˚C for 20 minutes. The solution was cooled to 80 ˚C and 1.2 mL Cs precursor solution was injected. The reaction proceeds for 90 seconds and was cooled in an ice bath. The resulting suspension was centrifuged at 7000 rpm for 10 minutes and the supernatant collected. The collected solution was allowed to sit for up to 24 hours and centrifuged whenever a significant amount of salt had precipitated. Once the solution remains clear, 1.5 the volume of acetone was added to precipitate the QDs. The QDs were dissolved in a minimum volume hexane and precipitated with methyl acetate. The resulting QDs were stored in hexane under Ar at 4 ˚C until use.

Preparation of $CsPbBr_3$ nanowires (NWs) and nanoplatelets (NPLs)
NWs and NPLs were prepared via methods recently developed by Dong et al.(*3*)For 2 nm-thick NWs, 120 mg $CoBr_2$, 65 mg $PbBr_2$ 5 mL ODE, 3 mL OA, and 3 mL OAm were loaded into a 25 mL three neck flask and vacuumed at room temperature for 10 minutes. For 2nm-thick NPLs, 400 mg $CuBr_2$, 100 mg $ZnBr_2$, 65 mg $PbBr_2$ 5 mL ODE, 3 mL OA, and 3 mL OAm were loaded into a 25 mL three neck flask and vacuumed at room temperature for 10 minutes. In both cases, the solution was vacuumed at 100 ˚C for 10 minutes then returned to Ar gas atmosphere. The solution was heated to 200 ˚C and held for 5 minutes to dissolve all the salt. The solution were cooled to room temperature (20-30 ˚C). 0.4 mL Cs precursor was cooled to room temperature in a syringe and injected into the solution and the solution as transferred to a centrifuge tube. 1.5 X the volume of acetone was added carefully so as not to directly mix the acetone and the precursor solution. The container was shaken vigorously to initiate the reaction and allowed to react for 2 minutes. The resulting suspension was centrifuged at 7000 rpm for 10 minutes. The supernatant was discarded, the particles were dissolved in 1mL hexane and centrifuged at 3000 rpm for 5 minutes to remove salts. The supernatant was collected and the particles were precipitated with 2mL methyl acetate and centrifuged at 7000 rpm for 10 minutes. The supernatant was discarded and the particles were dissolved in 2 mL hexane and centrifuged at 17000 rpm for 10 minutes, the supernatant collected and stored under Ar gas at 4 ˚C until used.

Ligand exchange of $CsPbBr_3$ NCs with dimethyl-dioctadecyl-ammonium bromide (DDAB)
10 µl NC stock solution was dried and dissolved in 200 µl toluene. ~10 µl of 0.05 M DDAB solution in toluene was added and the solution was stirred. The PL intensity was monitored until it reached maximum (the amount of DDAB can be adjusted batch to batch to reach maximum QY), the particles were precipitated with methyl acetate and centrifuged at 1700 rpm. The resulting nanocrystals were dispersed in 1% PS/toluene solution.

Preparation of NC film on sapphire substrate for temperature-dependent PL measurement
DDAB-passivated NCs were initially dispersed in 1% PS-toluene solution. PS polymer was used as the matrix to disperse the NCs in the dried film of NCs on the substrate. The concentration of NCs were adjusted such that the absorbance in 1 cm cell is approximately 5 near 400 nm. The solution was drop cast on a sapphire substrate and dried in a vacuum chamber for 30 minutes.

**S3. Sample Characterization**

Temperature-dependent photoluminescence measurements
NC film on a sapphire substrate were loaded into an open cycle cryostat (Oxford Instruments Microstat-HE), coupled to an inverted microscope (Nikon Ti-e), to record the photoluminescence in the temperature range of 5-300K. Excitation was provided with either 405 nm pulsed diode laser (Hamamatsu) with pulse width of 45 ps or 400 nm pulse obtained by frequency doubling 800 nm, 200 fs output from Ti:Sapphire laser (One-Five Origami, NKT Photonics). The intensity of excitation light was maintained below 0.2 $W/cm^2$ to minimize the heating of the sample by light absorption at cryogenic temperatures. Frequency doubled Ti:Sapphire laser was used for <20 ns time window on the streak camera, which gives maximum time resolution of ~20 ps. For all other measurements with >20 ns time window, pulsed diode laser that provides more flexible control of repetition rate was used as the excitation source.

Photoluminescence (PL) was collected using a 50X objective with a long working distance (15 mm) and sent to the imaging spectrograph EMCCD (Andor, iXon) for steady-state PL measurement. For time-resolved measurements, the PL was coupled to an imaging spectrograph to spectrally disperse the PL onto a Hamamatsu streak camera (C14831). Since the time resolution of the PL spectra depends on the total time window of the acquisition in the streak camera, several different time windows (1 ns, 20 ns, 100 ns, 2 μs and 20 μs) were used to cover the necessary time window to observe the dark exciton PL and obtain sufficient time resolution for the fast component of PL decay dynamics.

Magnetic field-dependent dark exciton PL lifetime measurements
For the measurement of magnetic field-dependent PL lifetime, the sample film on a sapphire substrate was mounted in a magneto optical cryostat oriented in Faraday geometry (Cryostat J2205). The NC sample was excited with 405nm, 150 ns excitation pulses produced at the repetition rate of 200 kHz using a CW laser and acousto-optic modulator. Time-resolved photoluminescence was collected by time-correlated single photon counting (PicoHarp 300) with an avalanche photodiode. 150 ns pulse width was sufficient to measure the PL decay dynamics of dark exciton with 1-10 μs lifetime.

Transmission Electron Microscope Measurements
TEM and STEM images were obtained with a double Cs-aberration corrected JEOL ARM-200F operated at 200 kV.

## S4. Temperature Dependent PL dynamics of different CsPbBr$_3$ NCs (Fig. S1-S5)

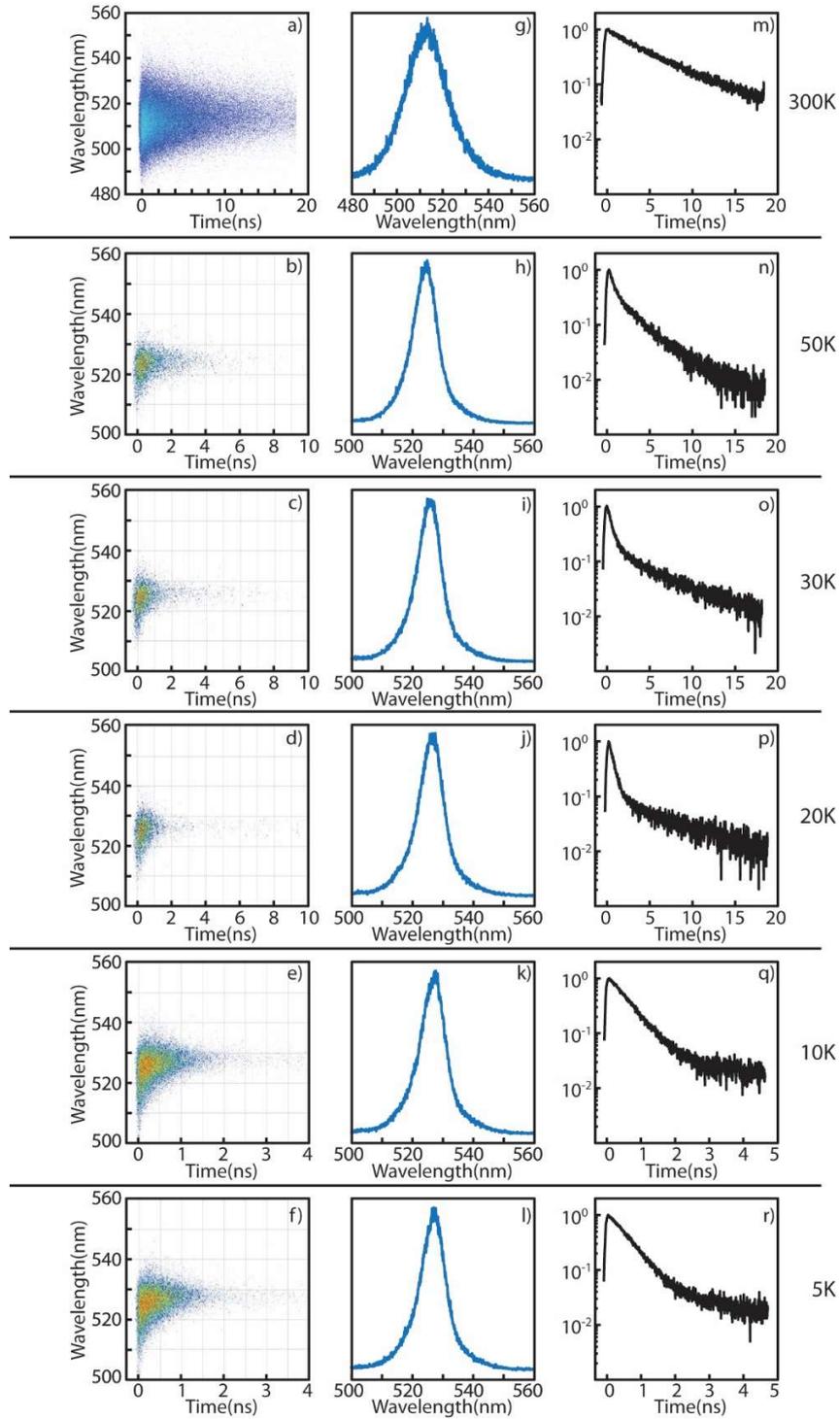

**Fig. S1**: Temperature-dependent PL spectra and PL decay dynamics of the weakly confined CsPbBr$_3$ NCs (10 nm size). Time-dependent PL spectra (a-f), time averaged PL spectra (g-l), and the spectrally averaged PL decay dynamics (m-r).

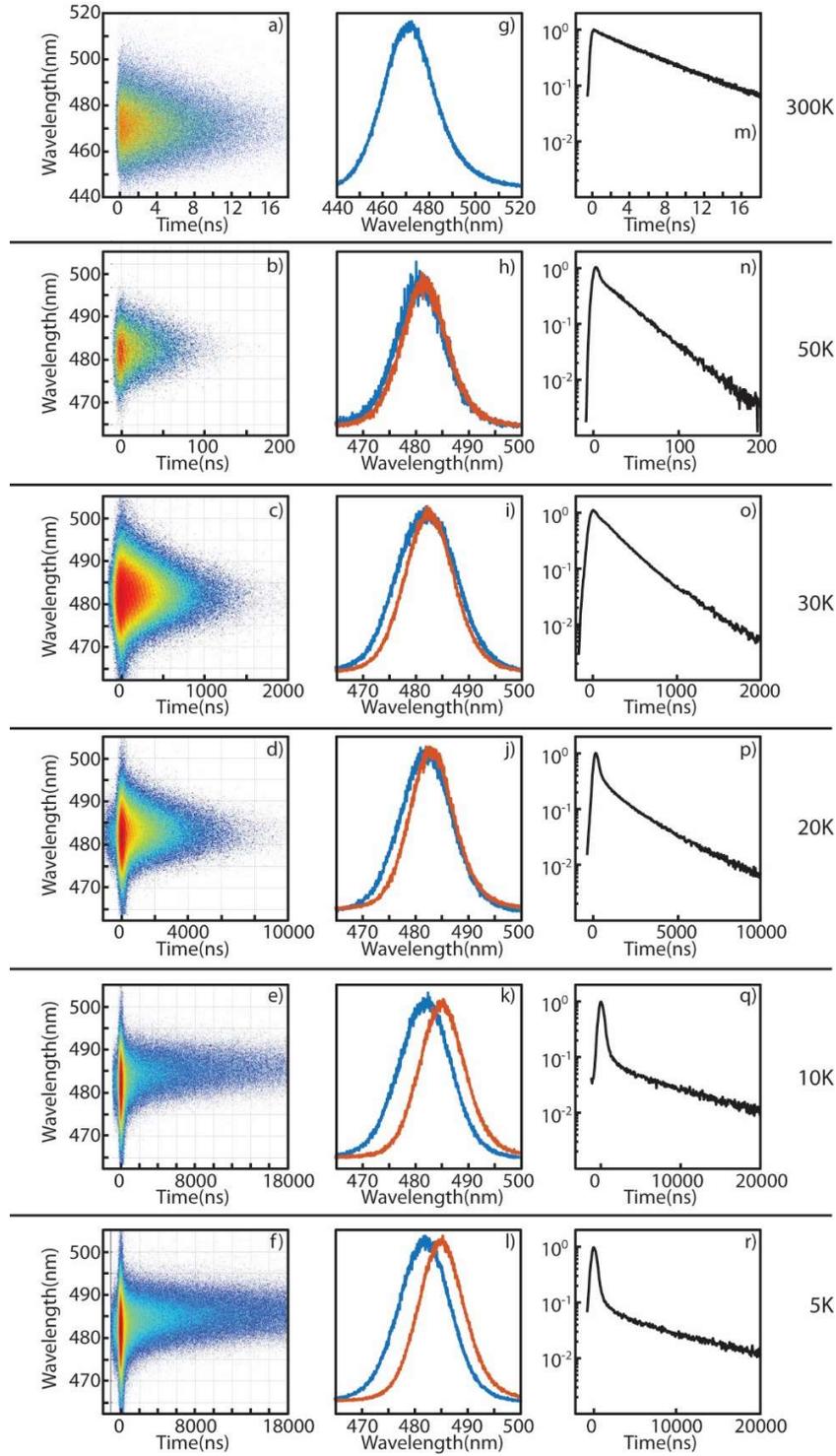

**Fig. S2**: Temperature-dependent PL spectra and PL decay dynamics of the strongly confined CsPbBr$_3$ QDs (4 nm size). Time-dependent PL spectra (a-f), time-gated PL spectra separating the spectra from the short (blue) and long (red) component of the decay (g-l), and the spectrally averaged PL decay kinetics (m-r).

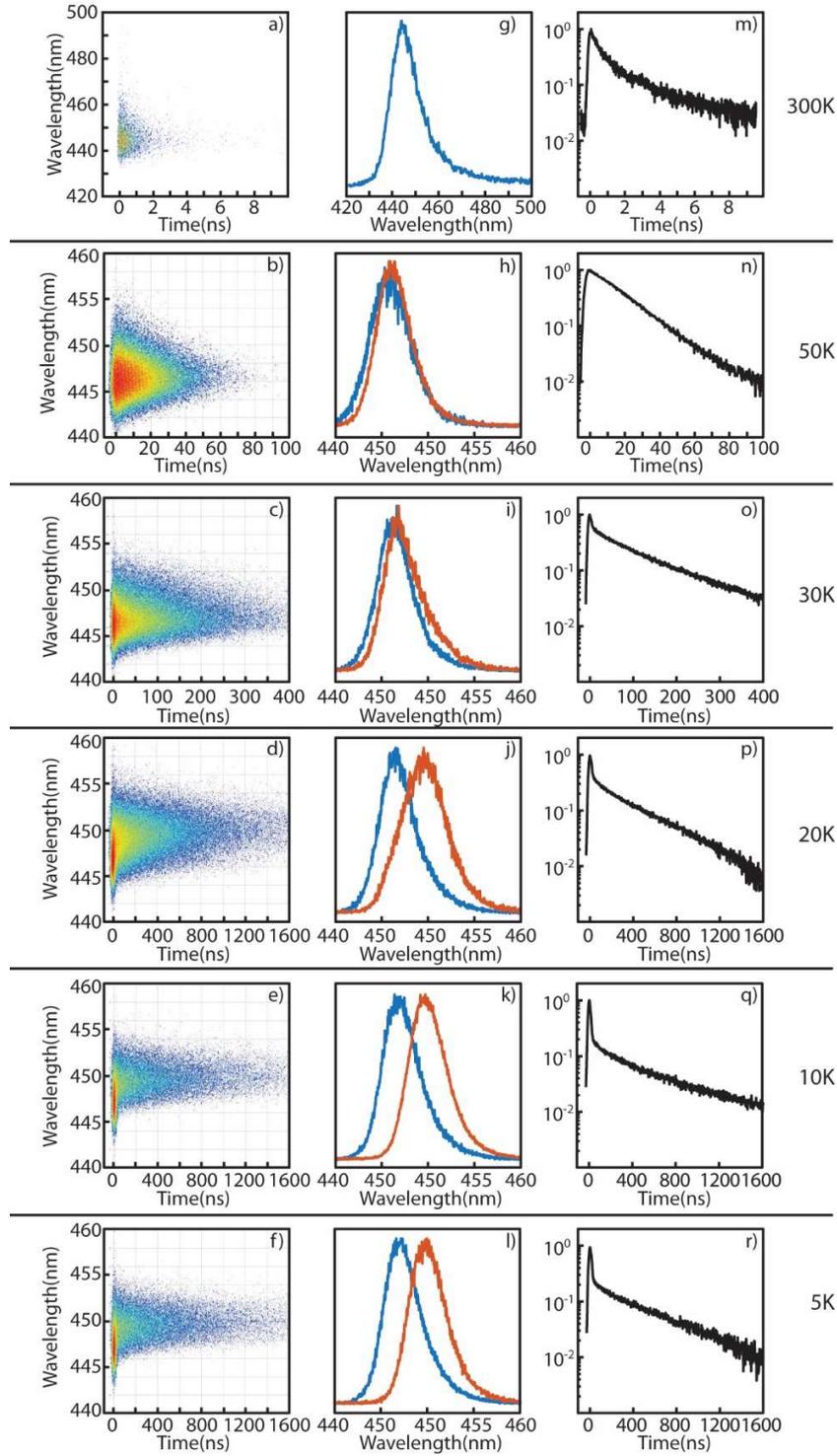

**Fig. S3**: Temperature-dependent PL spectra and PL decay dynamics of the strongly confined CsPbBr$_3$ NWs (2 nm thick). Time-dependent PL spectra (a-f), time-gated PL spectra separating the spectra from the short (blue) and long (red) component of the decay (g-l), and the spectrally averaged PL decay kinetics (m-r).

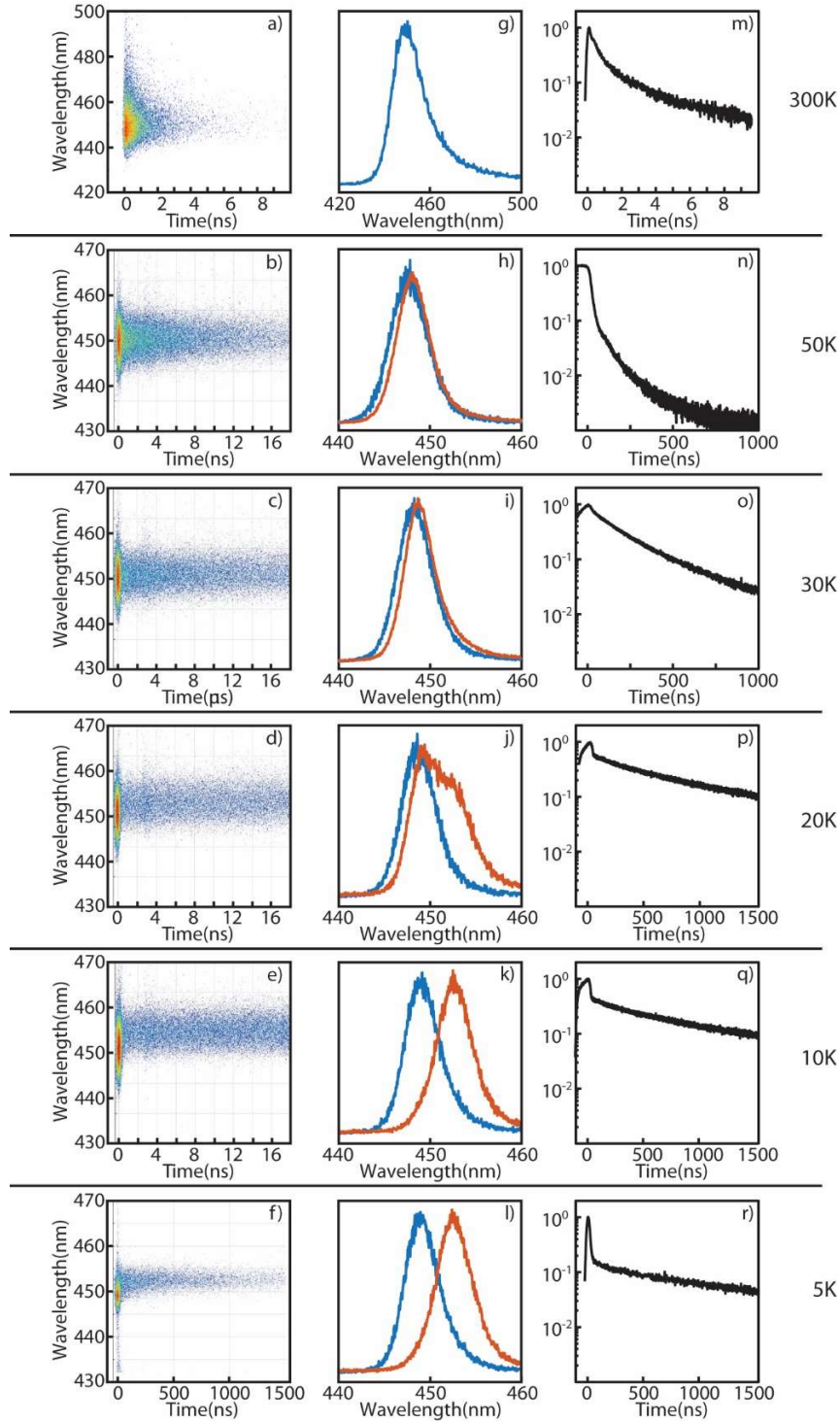

**Fig. S4:** Temperature-dependent PL spectra and PL decay dynamics of the strongly confined CsPbBr$_3$ NPLs (2 nm-thick). Time-dependent PL spectra (a-f), steady state PL spectra(g-l), and the spectrally averaged PL decay dynamics (m-r). The data shown in n-q were obtained using 150 ns-long excitation pulse, which shows artifactual signal at first early times.

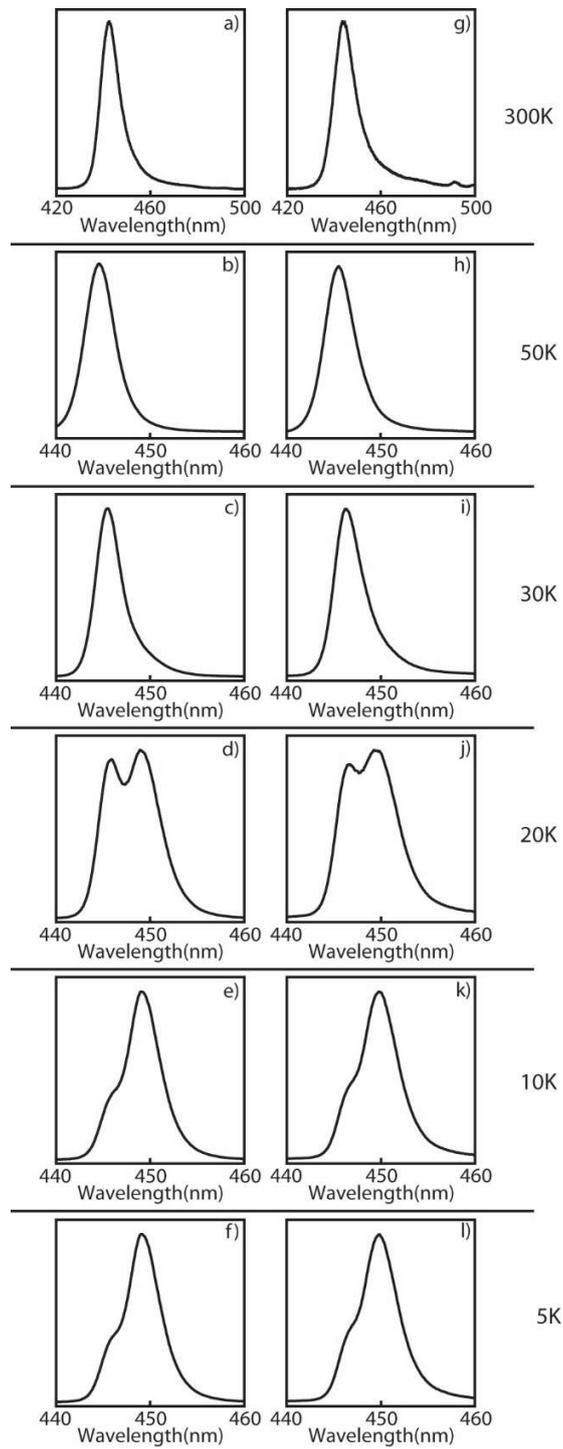

**Fig. S5:** Temperature-dependent steady state PL spectra of the strongly confined CsPbBr$_3$ NWs (a-f) and NPLs (g-l).

## S5. Size-dependence of $\Delta E_{BD}$ in Strongly Confined QDs and NPLs (Fig. S6-S7)

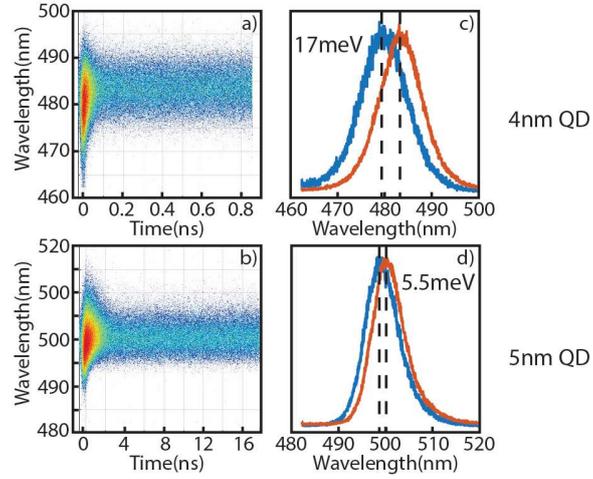

**Fig. S6:** Comparison of the time-dependent PL spectra (a,b) and time-gated PL spectra (c, d) of two sizes (4nm and 5nm) QDs at 5K. The size of the QDs are indicated next to each panel. Time gating was done near 0 ns and 10 ns to separate the bright and dark exciton PL spectra.

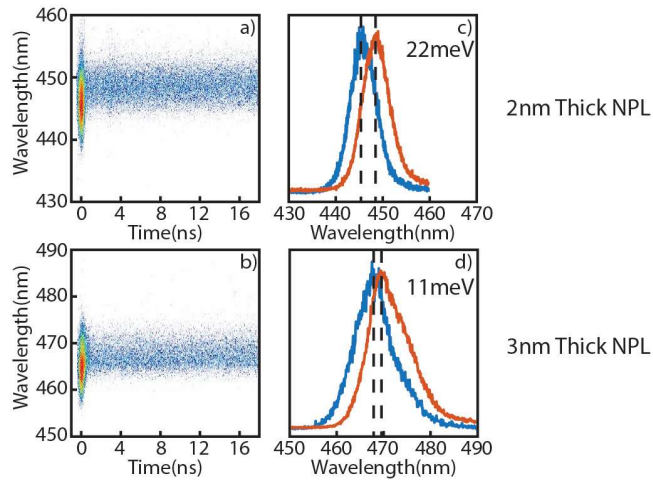

**Fig. S7:** Comparison of the time-dependent PL spectra (a,b) and time-gated PL spectra (c, d) of the strongly confined NPLs of two different thicknesses at 5K. The thickness of the NPLs are indicated next to each panel. Time gating was done near 0 ns and 10 ns to separate the bright and dark exciton PL spectra.

## S6. Estimation of $\tau_B$ and $\tau_{BD}$ from the Steady State PL Spectra (Fig. S8-S9)

Time constants for the relaxation of bright exciton ($\tau_B$) and transition from bright to dark exciton ($\tau_{BD}$) were estimated from the steady state PL spectra (Fig. 2) and the fast-decay kinetics data obtained from streak camera on narrow time window (Fig. S8) at 5K. We extract accurate $\tau_{fast}$ values listed in Table 1 from Fig. S8.

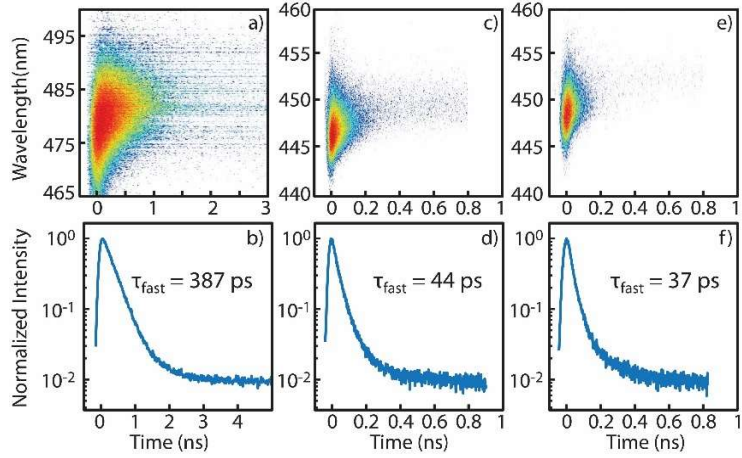

**Fig. S8:** PL decay dynamics from the QD (a,b), NW (c,d), and NPL (e,f) measured with a streak camera with narrow time window. Instrument response is sufficiently fast (~20 ps for 1 ns window) to reliably extract $\tau_{fast}$. The top row shows the time-resolved PL spectra and the bottom row shows the spectrally integrated PL decay kinetics with the long decay component subtracted.

At 5K, the fast-decay dynamics represent the sum of the relaxation of the bright exciton and transfer from bright to dark exciton level if dark exciton relaxation is sufficiently slow compared to all other processes. (See S4 for details of derivation)

$$1/\tau_B + 1/\tau_{BD} = 1/\tau_{fast} \quad (Eq. S1)$$

Because $\Delta E_{BD} \gg kT$ at 5K, thermal excitation from bright to dark state can be ignored. We can take the ratio of the PL intensities from bright ($I_B$) and dark ($I_D$) exciton in steady state PL spectra at 5K as the branching ratio for bright exciton relaxation and bright-to-dark transition under the following two assumptions: (i) The time scale for cooling from the initially excited level to the bandedge bright exciton level is much shorter than $\tau_B$ and $\tau_{BD}$, (ii) nonradiative decay of bright and dark exciton is not significant at 5 K.

$$\frac{\tau_B}{\tau_{BD}} = \frac{I_D}{I_B} \quad (Eq. S2)$$

Assumption (i) is reasonable since the time scale for cooling of hot exciton reported is typically sub ps in perovskite nanocrystals. Assumption (ii) is less straightforward, but we assume it is approximately valid considering the near-maximum PL intensity at 5 K as shown in Fig. S9.

**Sample calculation of τ_B and τ_BD for NPLs:**

Since the bright and dark excitons PL are spectrally well separated, we can obtain $\frac{I_D}{I_B}$ accurately from the fitting of the steady state PL shown in Fig. 2.

For the NPL sample, $\frac{I_D}{I_B} = 2.3$, therefore $\tau_B = 2.3\tau_{BD}$.

$1/\tau_B + 1/\tau_{BD} = 1/\tau_{fast} = \frac{1}{38\ ps}$ from Fig. S8 (f).

From the two equations above, we obtain $\tau_{BD} = 54\ ps$ and $\tau_B = 124\ ps$.

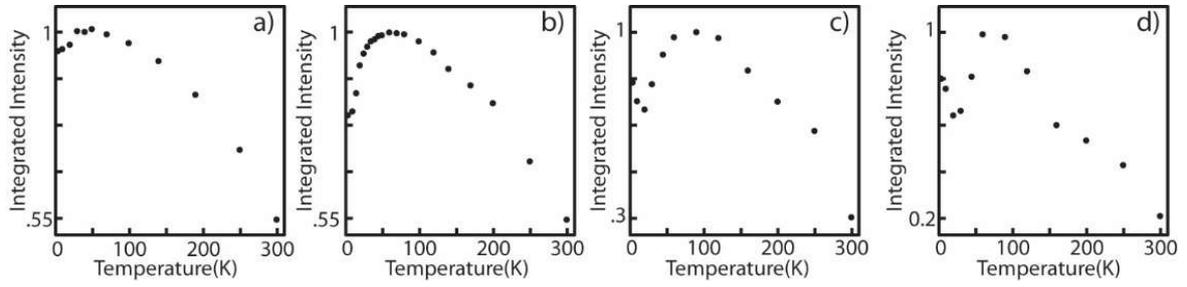

**Fig. S9:** Normalized temperature dependent integrated intensities from the weekly confined NC, and strongly confined QD, NW, and NPL films used in the main text(a-d).

## S7. Three-Level Kinetic Model and Temperature-Dependent τfast and τslow (Fig. S10)

It is common to model the kinetics of the bright and dark exciton after a three-level system with two emitting states (bright and dark) with lifetime $\tau_B$ and $\tau_D$ separated in energy by $\Delta E_{BD}$. Transition between these two states occurs via absorption or emission of a phonon with rates $\gamma_{DB} = \gamma_0 N_B$, and $\gamma_{BD} = \gamma_0(N_B + 1)$, where $\gamma_0$ is the zero temperature bright-to-dark transition rate and $N_B$ is the phonon occupation number. We can model the time dependent population dynamics of the bright ($P_B$) and dark ($P_D$) state as follows.

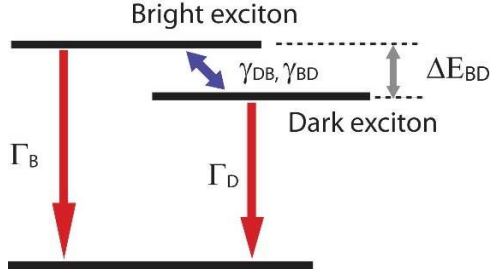

$$\frac{dP_B}{dt} = -\big(\Gamma_B + \gamma_0(N_B + 1)\big)P_B + \gamma_0 N_B P_D \quad \text{(Eq. S3)}$$

$$\frac{dP_D}{dt} = -(\Gamma_D + \gamma_0 N_B)P_D + \gamma_0(N_B + 1)P_B \quad \text{(Eq. S4)}$$

Where $\Gamma_B = \frac{1}{\tau_B}$, $\Gamma_D = \frac{1}{\tau_D}$, $\gamma_0(N_B + 1) = \frac{1}{\tau_{BD}}$, $\gamma_0 N_B = \frac{1}{\tau_{DB}}$, and $N_B = \frac{1}{(\exp(\Delta E_{BD}/kT)-1)}$ at temperature $T$.

Solving Eq. S3 and S4 with initial condition of $P_B(t=0) = 1$, $P_D(t=0) = 0$, we get the biexponential decay for both $P_B$(t) and $P_D$(t) with the two temperature-dependent rate constants $\Gamma_{slow}(T) = 1/\tau_{slow}(T)$ and $\Gamma_{fast}(T) = 1/\tau_{fast}(T)$ expressed as follows.

$$\Gamma_{(fast(+),slow(-))}(T) = \frac{1}{2}\Big(\Gamma_B + \Gamma_D + \gamma_0 + 2N_B\gamma_0 \pm \sqrt{(\Gamma_B - \Gamma_D + \gamma_0)^2 + 4N_B\gamma_0^2(N_B + 1)}\Big) \quad \text{(Eq. S5)}$$

At 5 K, $N_B \approx 0$. Under the condition $\Gamma_D \ll \Gamma_B$ and $\gamma_0$, we obtained the following two equations.

$$\Gamma_{slow}(5K) \approx \frac{1}{2}\Big(\Gamma_B + \Gamma_D + \gamma_0 - \sqrt{(\Gamma_B - \Gamma_D + \gamma_0)^2}\Big) = \Gamma_D = \frac{1}{\tau_D} \quad \text{(Eq. S6)}$$

$$\Gamma_{fast}(5K) \approx \frac{1}{2}\Big(\Gamma_B + \Gamma_D + \gamma_0 + \sqrt{(\Gamma_B - \Gamma_D + \gamma_0)^2}\Big) = \Gamma_B + \gamma_0 = \frac{1}{\tau_B} + \frac{1}{\tau_{BD}} \quad \text{(Eq. S7)}$$

Therefore, $\Gamma_D$, $\Gamma_B$ and $\gamma_0$ can be experimentally determined from PL decay kinetics at 5K and $\Gamma_{slow}(T)$ and $\Gamma_{fast}(T)$ can be obtained from Eq. S5 using these parameters.

Fig. S10 (a-c) shows the comparison of the experimental PL decay dynamics at 5K and the calculated time-dependent PL intensity, *I(t)*, at 5K obtained using Eq. S3, S4 and S8. The time-dependent bright and dark exciton populations are converted to the time dependent PL intensity, *I(t)*, using Eq. S8(*4*), where the calculated PL decay kinetics is convoluted with the instrument response function (IRF):

$$I(t) = IRF \otimes [\Phi_B \Gamma_B P_B(t) + \Phi_D \Gamma_D P_D(t)] \quad (Eq. S8)$$

$\Phi_B$ and $\Phi_D$ are the bright and dark exciton quantum yield (assumed to be the same). Fig. S10 (d-f) shows the comparison of the result from Eq. S5 with the experimentally measured temperature dependent $\tau_{slow}$ and $\tau_{fast}$ for strongly confined QDs, NWs and NPLs as nonadjustable parameters. The model fits the experimental data for QDs well. On the other hand, the model overestimates τ$_{fast}$ of NWs and NPLs, suggesting the presence of an additional thermally activated decay pathway of bright exciton consistent with the more complicated change in relative PL QY for the NWs and NPLs as shown in Fig. S10.

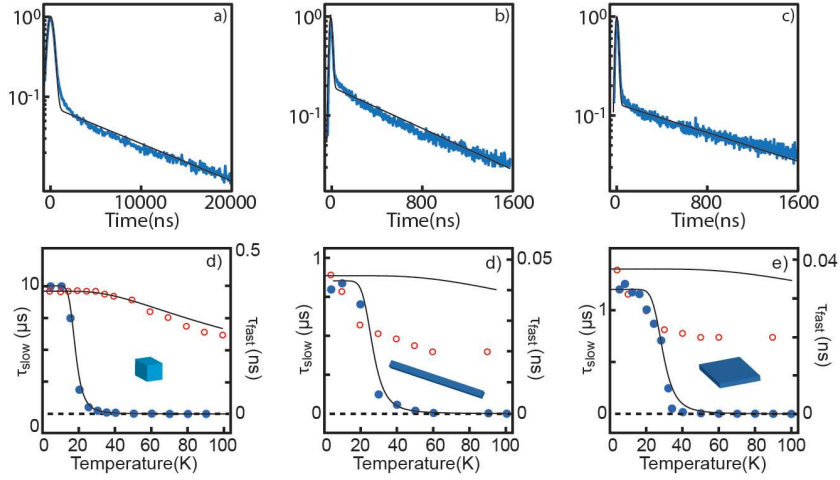

**Fig. S10:** (a-c) Comparison of experimental (blue) and calculated (black) time-dependent PL intensity for QDs (a), NWs (b) and NPLs (c). (d-f) Temperature dependent lifetime data fit with equation S6 and S7.